\documentclass[aps,prl,english,twocolumn,superscriptaddress,showpacs,floats]{revtex4}
\usepackage{epsfig,epsf,amsmath,amsthm,amsfonts,
            graphics,graphicx,amssymb,babel}

\input psfig.sty

\newcommand{\eqlum}{\cal{L}}

\begin{document}
\lefthyphenmin=2
\righthyphenmin=3

%
\title{
Observation of Exclusive Electron-Positron Production in Hadron-Hadron Collisions
}

\affiliation{Institute of Physics, Academia Sinica, Taipei, Taiwan 11529, Republic of China} 
\affiliation{Argonne National Laboratory, Argonne, Illinois 60439} 
\affiliation{Institut de Fisica d'Altes Energies, Universitat Autonoma de Barcelona, E-08193, Bellaterra (Barcelona), Spain} 
\affiliation{Baylor University, Waco, Texas  76798} 
\affiliation{Istituto Nazionale di Fisica Nucleare, University of Bologna, I-40127 Bologna, Italy} 
\affiliation{Brandeis University, Waltham, Massachusetts 02254} 
\affiliation{University of California, Davis, Davis, California  95616} 
\affiliation{University of California, Los Angeles, Los Angeles, California  90024} 
\affiliation{University of California, San Diego, La Jolla, California  92093} 
\affiliation{University of California, Santa Barbara, Santa Barbara, California 93106} 
\affiliation{Instituto de Fisica de Cantabria, CSIC-University of Cantabria, 39005 Santander, Spain} 
\affiliation{Carnegie Mellon University, Pittsburgh, PA  15213} 
\affiliation{Enrico Fermi Institute, University of Chicago, Chicago, Illinois 60637} 
\affiliation{Joint Institute for Nuclear Research, RU-141980 Dubna, Russia} 
\affiliation{Duke University, Durham, North Carolina  27708} 
\affiliation{Fermi National Accelerator Laboratory, Batavia, Illinois 60510} 
\affiliation{University of Florida, Gainesville, Florida  32611} 
\affiliation{Laboratori Nazionali di Frascati, Istituto Nazionale di Fisica Nucleare, I-00044 Frascati, Italy} 
\affiliation{University of Geneva, CH-1211 Geneva 4, Switzerland} 
\affiliation{Glasgow University, Glasgow G12 8QQ, United Kingdom} 
\affiliation{Harvard University, Cambridge, Massachusetts 02138} 
\affiliation{Division of High Energy Physics, Department of Physics, University of Helsinki and Helsinki Institute of Physics, FIN-00014, Helsinki, Finland} 
\affiliation{University of Illinois, Urbana, Illinois 61801} 
\affiliation{The Johns Hopkins University, Baltimore, Maryland 21218} 
\affiliation{Institut f\"{u}r Experimentelle Kernphysik, Universit\"{a}t Karlsruhe, 76128 Karlsruhe, Germany} 
\affiliation{High Energy Accelerator Research Organization (KEK), Tsukuba, Ibaraki 305, Japan} 
\affiliation{Center for High Energy Physics: Kyungpook National University, Taegu 702-701, Korea; Seoul National University, Seoul 151-742, Korea; and SungKyunKwan University, Suwon 440-746, Korea} 
\affiliation{Ernest Orlando Lawrence Berkeley National Laboratory, Berkeley, California 94720} 
\affiliation{University of Liverpool, Liverpool L69 7ZE, United Kingdom} 
\affiliation{University College London, London WC1E 6BT, United Kingdom} 
\affiliation{Centro de Investigaciones Energeticas Medioambientales y Tecnologicas, E-28040 Madrid, Spain} 
\affiliation{Massachusetts Institute of Technology, Cambridge, Massachusetts  02139} 
\affiliation{Institute of Particle Physics: University of Alberta, Edmonton, Canada T6G~2G7; McGill University, Montr\'{e}al, Canada H3A~2T8; and University of Toronto, Toronto, Canada M5S~1A7} 
\affiliation{University of Michigan, Ann Arbor, Michigan 48109} 
\affiliation{Michigan State University, East Lansing, Michigan  48824} 
\affiliation{Institution for Theoretical and Experimental Physics, ITEP, Moscow 117259, Russia} 
\affiliation{University of New Mexico, Albuquerque, New Mexico 87131} 
\affiliation{Northwestern University, Evanston, Illinois  60208} 
\affiliation{The Ohio State University, Columbus, Ohio  43210} 
\affiliation{Okayama University, Okayama 700-8530, Japan} 
\affiliation{Osaka City University, Osaka 588, Japan} 
\affiliation{University of Oxford, Oxford OX1 3RH, United Kingdom} 
\affiliation{University of Padova, Istituto Nazionale di Fisica Nucleare, Sezione di Padova-Trento, I-35131 Padova, Italy} 
\affiliation{LPNHE, Universite Pierre et Marie Curie/IN2P3-CNRS, UMR7585, Paris, F-75252 France} 
\affiliation{University of Pennsylvania, Philadelphia, Pennsylvania 19104} 
\affiliation{Istituto Nazionale di Fisica Nucleare Pisa, Universities of Pisa, Siena and Scuola Normale Superiore, I-56127 Pisa, Italy} 
\affiliation{University of Pittsburgh, Pittsburgh, Pennsylvania 15260} 
\affiliation{Purdue University, West Lafayette, Indiana 47907} 
\affiliation{University of Rochester, Rochester, New York 14627} 
\affiliation{The Rockefeller University, New York, New York 10021} 
\affiliation{Istituto Nazionale di Fisica Nucleare, Sezione di Roma 1, University of Rome ``La Sapienza," I-00185 Roma, Italy} 
\affiliation{Rutgers University, Piscataway, New Jersey 08855} 
\affiliation{Texas A\&M University, College Station, Texas 77843} 
\affiliation{Istituto Nazionale di Fisica Nucleare, University of Trieste/\ Udine, Italy} 
\affiliation{University of Tsukuba, Tsukuba, Ibaraki 305, Japan} 
\affiliation{Tufts University, Medford, Massachusetts 02155} 
\affiliation{Waseda University, Tokyo 169, Japan} 
\affiliation{Wayne State University, Detroit, Michigan  48201} 
\affiliation{University of Wisconsin, Madison, Wisconsin 53706} 
\affiliation{Yale University, New Haven, Connecticut 06520} 
\author{A.~Abulencia}
\affiliation{University of Illinois, Urbana, Illinois 61801}
\author{J.~Adelman}
\affiliation{Enrico Fermi Institute, University of Chicago, Chicago, Illinois 60637}
\author{T.~Affolder}
\affiliation{University of California, Santa Barbara, Santa Barbara, California 93106}
\author{T.~Akimoto}
\affiliation{University of Tsukuba, Tsukuba, Ibaraki 305, Japan}
\author{M.G.~Albrow}
\affiliation{Fermi National Accelerator Laboratory, Batavia, Illinois 60510}
\author{D.~Ambrose}
\affiliation{Fermi National Accelerator Laboratory, Batavia, Illinois 60510}
\author{S.~Amerio}
\affiliation{University of Padova, Istituto Nazionale di Fisica Nucleare, Sezione di Padova-Trento, I-35131 Padova, Italy}
\author{D.~Amidei}
\affiliation{University of Michigan, Ann Arbor, Michigan 48109}
\author{A.~Anastassov}
\affiliation{Rutgers University, Piscataway, New Jersey 08855}
\author{K.~Anikeev}
\affiliation{Fermi National Accelerator Laboratory, Batavia, Illinois 60510}
% ...
\collaboration{CDF Collaboration}
\noaffiliation

%
%
\begin{abstract}
  We present the first observation of exclusive $e^+e^-$ production in hadron-hadron collisions, using $p\bar{p}$ collision data at \mbox{$\sqrt{s}=1.96$~TeV} taken by the Run II Collider Detector at Fermilab, and corresponding  to an integrated luminosity of \mbox{$532$~pb$^{-1}$}.  We require the absence of any particle signatures in the detector except for an electron and a positron candidate, each with transverse energy \mbox{$E_T>5$~GeV} and pseudorapidity \mbox{$|\eta|<2$}.   With these criteria, $16$ events are observed compared to a background expectation of \mbox{$1.9\pm0.3$} events.  These events are consistent in cross section and properties with the QED process \mbox{$p\bar{p} \rightarrow p + e^+e^- + \bar{p}$}  through two-photon exchange.  The measured cross section is \mbox{$1.6^{+0.5}_{-0.3}\mathrm{(stat)}\pm0.3\mathrm{(syst)}$~pb}.  This agrees with the theoretical prediction of \mbox{$1.71 \pm 0.01$~pb}.   

\end{abstract}

\maketitle


In this Letter we report an observation of exclusive dilepton
production in hadron-hadron collisions.  In an exclusive dilepton
process there are no particles produced other than the lepton pair and
the incident hadrons do not dissociate.  The process
\mbox{$\gamma\gamma \rightarrow \ell^+\ell^-$} in hadron collisions is
discussed in Ref.~\cite{bhabhabudnev}.  In this Letter we consider
\mbox{$p\bar{p} \rightarrow p + e^+e^- + \bar{p}$} where the electron
pair is produced via \mbox{$\gamma\gamma \rightarrow e^+e^-$}.  The
events are simulated using the {\sc lpair} Monte Carlo
generator~\cite{vermaseren}.

Exclusive events have been proposed as a search channel for new
physics at the Large Hadron Collider (LHC)~\cite{newphyslhc} because,
if the momentum of the outgoing protons is measured, exclusive
processes provide an additional method to calculate the invariant mass
of the central system ~\cite{fp420}.  Also, since the interaction
depends only on quantum electrodynamics and the proton electromagnetic
form factor at small momentum transfer, the cross section is known
with an accuracy better than $1\%$.  This makes exclusive dilepton
processes potentially interesting candidates for improving the typical
$5\%$ uncertainty on the luminosity measurement at hadron
colliders~\cite{lhclumi}.  Exclusive dilepton events can provide an
excellent control sample for other exclusive processes whose
theoretical predictions are less certain, such as exclusive diphoton
production~\cite{cdfgg} and exclusive Higgs
production~\cite{newphyslhc}.  At the LHC, exclusive dilepton events
may also be used to calibrate forward proton
spectrometers~\cite{fp420}.  Two-photon production of lepton pairs has
been experimentally observed in $ep$~\cite{ep}, and $AA$~\cite{aa}
collisions, but the exclusive process has never been observed in
hadron-hadron collisions.  There were indications of exclusive muon
pair production at the Intersecting Storage Rings~\cite{isr}
(unpublished).

The present analysis uses a data set corresponding to an integrated
luminosity ${\cal L} = 532 \pm 32$~pb$^{-1}$ collected by the Run II
Collider Detector at Fermilab (CDF II)~\cite{CDF}.  CDF II is a
general purpose particle detector at the Tevatron, a $p\bar{p}$
collider with center of mass energy \mbox{$\sqrt{s}=1.96$~TeV}.  The
experimental signature of exclusive $e^+e^-$ production is an $e^+e^-$
pair with no other particles detected. The proton and antiproton lose
a small fraction of their momentum in these collisions and escape
along the beam direction without being detected.

We briefly describe the components of CDF II that are used in this
analysis~\cite{geometry}.  Both the central calorimeter~\cite{ccal},
covering $|\eta|<1.1$, and the plug calorimeter~\cite{pem}, covering
\mbox{$1.1<|\eta|<3.6$}, have separate electromagnetic
(lead-scintillator) and hadronic (steel-scintillator)
compartments. The energy resolution for the central electromagnetic
calorimeter is $\sigma(E)/\sqrt{E}=13.5\% \oplus 1.5\%$, and for the
plug electromagnetic calorimeter it is $\sigma(E)/\sqrt{E}=16\% \oplus
1\%$.  The central electromagnetic calorimeters contain strip chambers
that measure the transverse profile of the electromagnetic shower at a
depth of six radiation lengths.  The electromagnetic calorimeters are
also capable of determining the arrival time of a shower in the
calorimeter with a resolution of $0.6$~ns~\cite{emtiming}.  The
miniplug calorimeter~\cite{miniplug} is a lead-liquid scintillator
sampling calorimeter and covers the region \mbox{$3.6<|\eta|<5.2$}.
Covering \mbox{$5.4<|\eta|<7.4$} is a series of three scintillator
paddles on each side of the detector, collectively called the beam
shower counters (BSC), labeled BSC1, BSC2, and BSC3, in order of
increasing $|\eta|$.  BSC1 is preceded by $2$ radiation lengths of
lead.

Inside the calorimeter is a tracking system composed of a solenoid
producing a 1.4~T magnetic field, the silicon tracking detector, and
the central outer tracker (COT)~\cite{COT}.  The tracking system
measures charged particle tracks in \mbox{$|\eta|<1.0$} with an
efficiency near unity, and charged particle tracks in
\mbox{$1.0<|\eta|<2.8$} with reduced efficiency.

The data were collected using a trigger requiring two
clusters~\cite{cluster} of energy in the central or plug
electromagnetic calorimeters and no activity in the BSC1 counters.
Events containing two electron candidates with $E_T>5$~GeV and
$|\eta|<2.0$ are selected.  Here, and in the remainder of this Letter,
we use the term ``electron'' to mean either electron or positron
candidate.  Electron candidates are defined as clusters in the
electromagnetic calorimeter with a hadronic-to-electromagnetic energy
ratio and lateral shower shape consistent with an electron, and a
track with \mbox{$p_T>1$~GeV/$c$} pointing to the calorimeter cluster.
The efficiency for triggering, reconstructing, and identifying both
electrons in a signal event is \mbox{$\varepsilon_{\mathrm{ee}} =
  (26\pm3)\%$}.  A large contribution to $\varepsilon_{\mathrm{ee}}$
comes from the electron trigger efficiency, ($77\pm5$)\% per electron,
due to the trigger turn-on at the threshold.

Events caused by cosmic ray interactions occur at random times, so we
require both clusters to have a measurement in the electromagnetic
calorimeter timing system with times that are consistent with the
collision.  The overall efficiency for the cosmic rejection
requirement is $\varepsilon_{\rm{cosmic}} = (93\pm3)\%$. 

To select exclusive events, we require that no additional particle
signatures be detected in the calorimeters or BSC.  A particle
signature in the calorimeter is defined as a cluster of adjacent
towers above the noise threshold in the miniplug, or a single tower
above the noise threshold in the plug and central calorimeters.  An
additional cluster is defined as a cluster that is not part of either
electron in the event.  A particle signature in the BSC is defined as
any hit above the noise threshold of the BSC.

When an inelastic interaction occurs in the same bunch crossing as an
exclusive event, it will deposit energy in the calorimeters and BSC,
preventing the identification of the exclusive event.  This efficiency
$\varepsilon_{\mathrm{exc}}$ is the fraction of zero-bias events
(triggered solely on the bunch crossing time) that pass the
exclusivity selection divided by the total number of zero-bias events.
This is directly related to the instantaneous bunch
luminosity~\cite{bunchlumi}.  Using the same run range for the
zero-bias data sample as the exclusive $e^+e^-$ triggered events, the
overall exclusive efficiency is \mbox{$\varepsilon_{\mathrm{exc}} =
  8.6\%$}, with negligible statistical uncertainty.  The systematic
uncertainty on this fraction is negligible compared to the uncertainty
of the luminosity.

As a consequence of the exclusivity selection requirements, electrons
that emit sufficient brem\-sstrah\-lung (final state radiation, FSR)
to deposit energy outside the electron's cluster will be excluded from
the candidate sample.  The corresponding efficiency is calculated by
generating events with the {\sc lpair} simulation plus the {\sc
  geant}-based~\cite{geant} detector simulation~\cite{cdfsim}, then
requiring the events to pass the exclusivity selection requirements.
This efficiency is \mbox{$\varepsilon_{\mathrm{FSR}} = (79\pm5)\%$}.
The dominant source of uncertainty in $\varepsilon_{\mathrm{FSR}}$ is
the material count in the detector.

The overall efficiency to identify exclusive $e^+e^-$ events with the
luminosity distribution of this data sample is $(1.6\pm0.2)\%$.  This
is dominated by $\varepsilon_{\mathrm{exc}}$, which is low because of
the large fraction of bunch crossings with more than one inelastic
$p\bar{p}$ interaction.

\begin{figure}[!h]
  \begin{center}
    \includegraphics[width=0.45\textwidth]{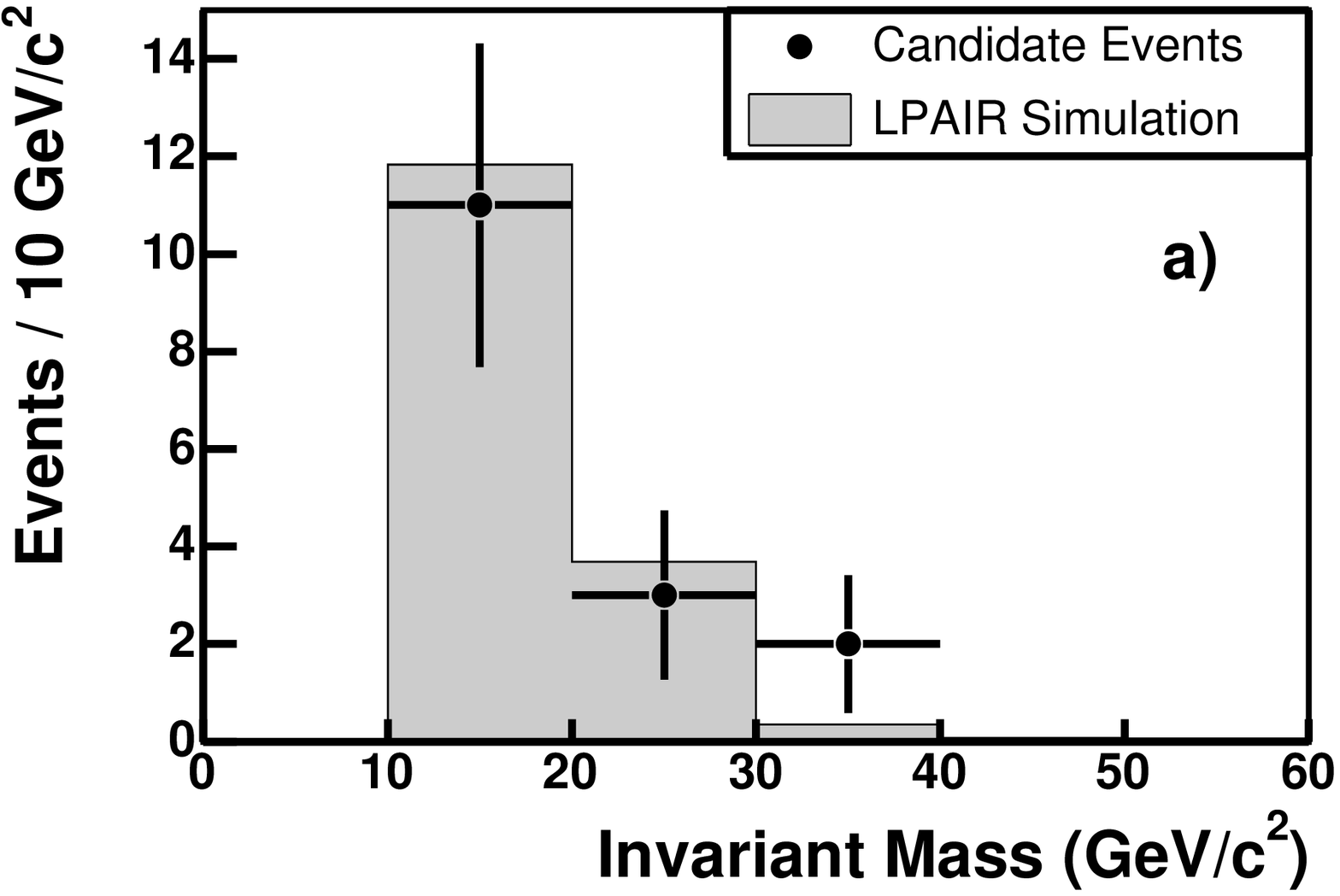}
    \includegraphics[width=0.45\textwidth]{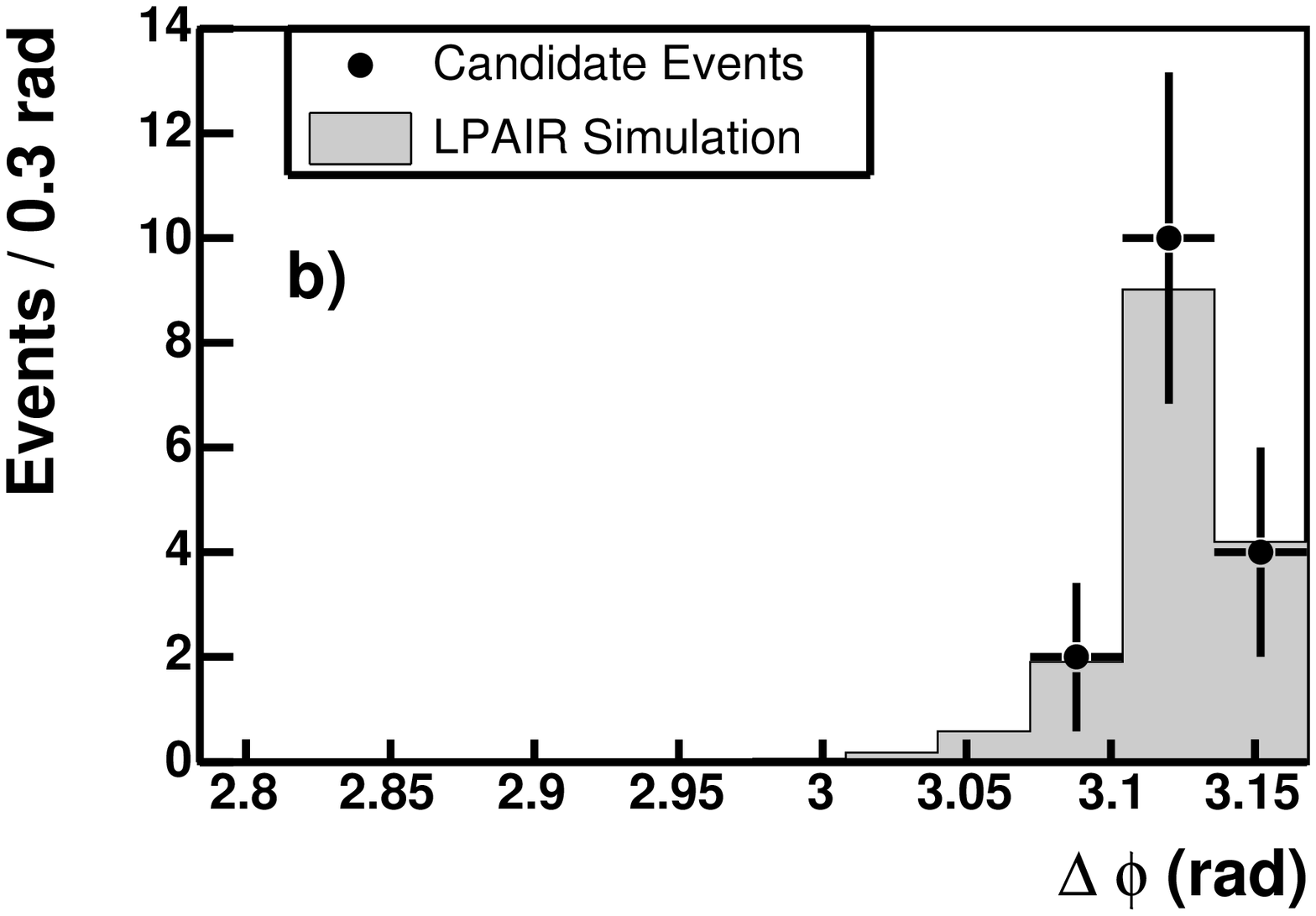}
    \caption{The 16 candidate events (points) compared to the {\sc lpair} simulation (solid). a) invariant mass of $e^+e^-$ pair (resolution $\sim$8\%), b) $\Delta\phi$ between $e^+e^-$ pair. The simulation is normalized to the number of data events.\label{fig:datavsmc} }
  \end{center}
\end{figure}

A total of 16 candidate events pass electron identification, cosmic
ray rejection, and exclusivity requirements.  All of these events
contain oppositely-charged electron pairs.  The invariant mass and
$\Delta \phi$ (the opening angle between the $e^+$ and $e^-$ in
$r$-$\phi$ plane) distributions of the candidate sample are compared
to events generated with {\sc lpair} plus detector simulation in
Fig.~\ref{fig:datavsmc}.   Other variables, such as the $p_T$ of the
$e^+e^-$ pair and $\eta$ of the electrons are also in agreement with
{\sc lpair}.

There are four backgrounds to consider: jets that pass electron
requirements (jet fakes), cosmic rays that interact in the detector,
non-exclusive events, and \mbox{$\gamma\gamma \rightarrow e^+e^-$}
events with proton dissociation.

The jet fake background consists of exclusive events with a hadronic
final state, such as $\pi^+\pi^-$, that fake a signal event when both
hadrons are reconstructed in the detector as electrons. In a
jet-triggered sample we observe no events that pass the exclusivity
cuts but fail the electron requirements. This results in an upper
limit on the rate of this process. Multiplying this upper limit with
the measured probability of a generic calorimeter cluster to pass the
electron cuts, an upper limit on the jet fake background is determined
to be 0.1 events.  Thus the jet fake background is taken to be
\mbox{$0.0^{+0.1}_{-0.0}$} events.

By examining the rate of out-of-time cosmic ray events, the
probability that a candidate event originates from a cosmic ray is
found to be negligible.

\begin{figure}[htb]
  \begin{center}
     \includegraphics[width=0.45\textwidth]{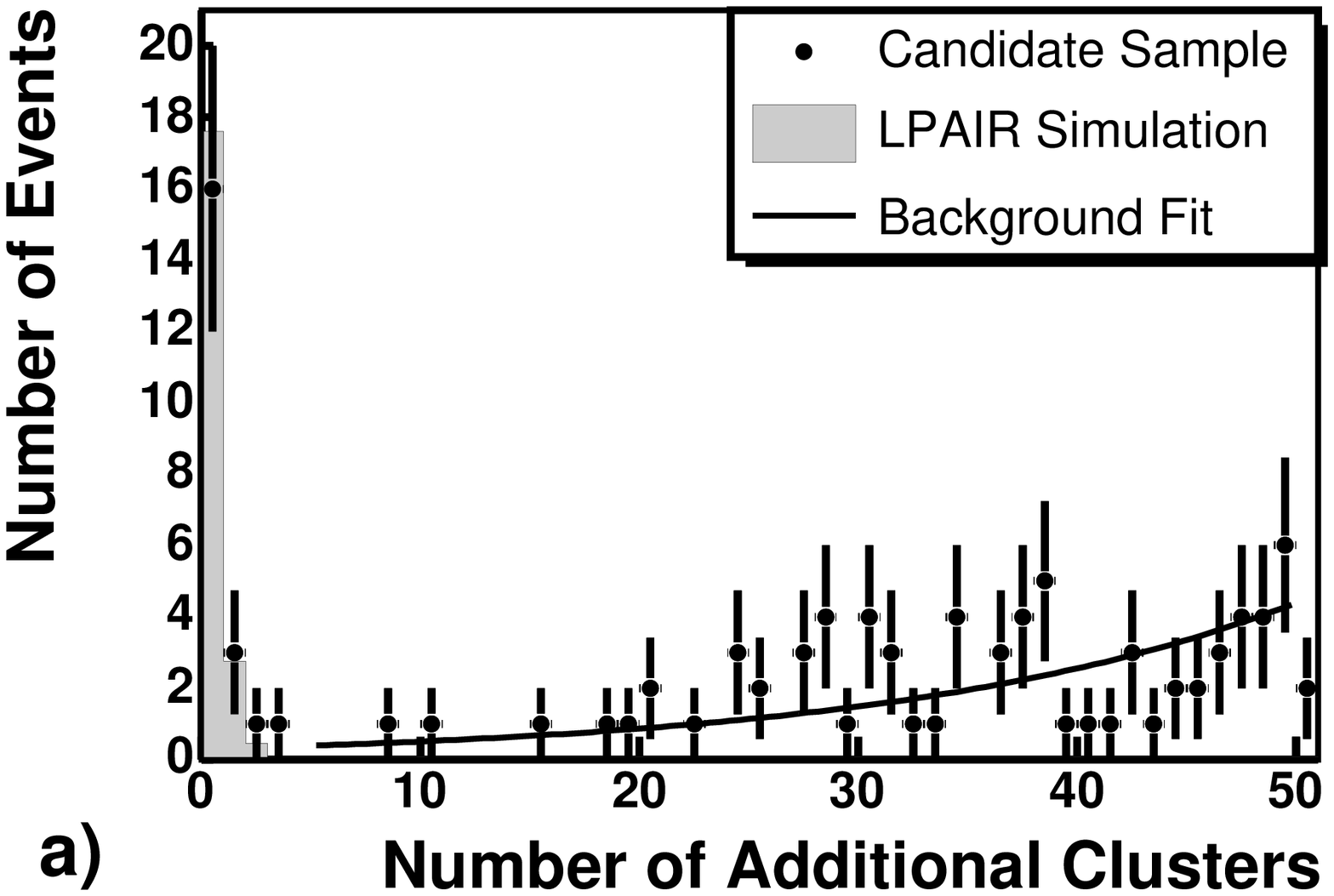}
     \includegraphics[width=0.45\textwidth]{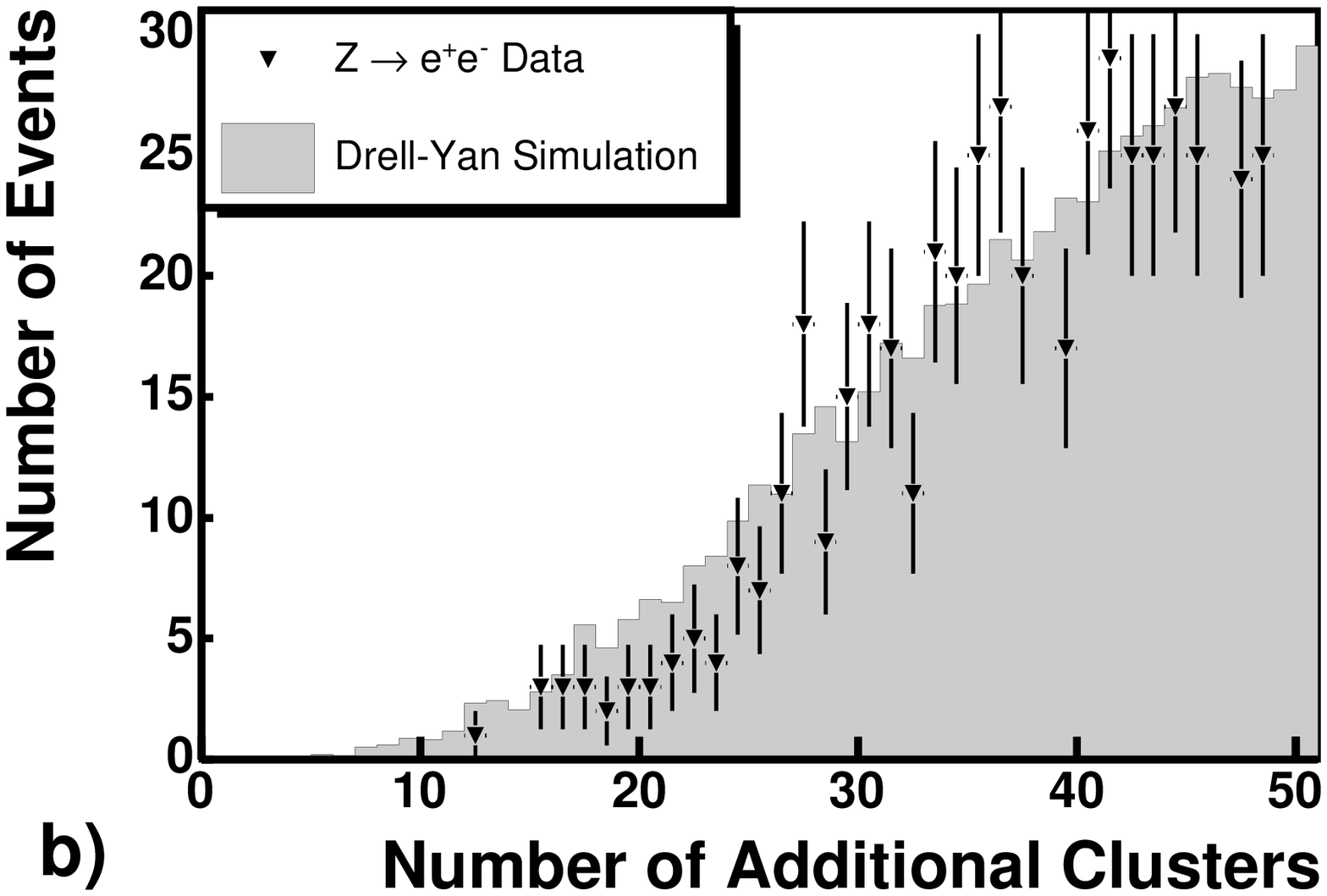}
    \caption{a) The number of additional clusters in {\sc lpair}
      simulation and signal sample events with all selection criteria applied
      except the exclusivity requirement.  {\sc lpair} simulation events are
      normalized to the number of data events with less than 5 clusters.  The line shows
      the fit used for the background estimate.  b) The number of
      additional clusters for $Z\rightarrow e^+e^-$ data and
      Drell-Yan simulation, the simulation is normalized to the number of data events. \label{fig:excBkg_zee}}
  \end{center}
\end{figure}

The third background is due to non-exclusive events in which one or
more particles passed undetected through cracks in the calorimeter, or
were below the noise thresholds, causing the event to appear
exclusive. The multiplicity of additional clusters is shown in
Figure~\ref{fig:excBkg_zee}a).  A clear peak is observed in the zero
additional clusters signal region.  To estimate the amount of
background in the signal region, the events with 5 to 50 additional
clusters were fit to the function $y=e^{(ax+b)}$, where the result of
the fit yields $a=0.05\pm0.01$ and $b=-1.3\pm0.4$.  The fit result
predicts a non-exclusive background of $0.3\pm0.1$ events with zero
additional clusters.

To verify that a peak in the signal region is not expected with
inclusive events we plot the number of associated clusters for a
sample of \mbox{$Z \rightarrow e^+e^-$} events selected from the same
run range as the exclusive $e^+e^-$ data set (the $Z$ boson cannot be
produced exclusively at a hadron collider due to Yang's
Theorem~\cite{yang}) as well as a sample of inclusive Drell-Yan
events, simulated with the {\sc pythia}~\cite{pythia} Monte Carlo
generator, in Fig.~\ref{fig:excBkg_zee}b).  The \mbox{$Z \rightarrow
  e^+e^-$} data sample contains events with two electrons that have an
invariant mass between 81~GeV/$c^2$ and 101~GeV/$c^2$ and a single
vertex reconstructed in the event.  There is no peak at zero
additional clusters in either the \mbox{$Z \rightarrow e^+e^-$} sample
nor the Drell-Yan simulated sample.  We also verified that this
distribution is nearly independent of the dielectron invariant mass
value using the simulated Drell-Yan sample.

The dissociation background arises from events that are mediated by
two-photon exchange, but instead of being truly exclusive, one or both
protons are excited into a low-mass state that dissociates. There is a
small probability that these dissociations will escape detection in
the BSC, and hence would not be distinguished from exclusive events.
We use two simulation programs, {\sc grape-dilepton}~\cite{grape} and
{\sc lpair}, to estimate this background.

The {\sc lpair} ({\sc grape-dilepton}) simulation predicts that 7\%
(5\%) of proton dissociations will have all dissociation products too
far forward to be detectable.  The average of the two estimates,
\mbox{$(6\pm1)\%$}, is taken as the probability that a dissociated
proton will escape detection.  To estimate the number of dissociation
background events in the candidate sample, the cross sections for
events in which one or both protons dissociate are extracted from {\sc
  lpair} and then multiplied by the probability that the dissociation
will escape detection.  The number of dissociation events in the 16
event signal sample is estimated to be $1.6\pm0.3$.  Therefore, the
sum of all background sources is \mbox{$N_{\mathrm{bkgd}}=1.9\pm0.3$}
events.

The cross section $\sigma$ for exclusive $e^+e^-$ production with $E_T
> 5$~GeV and $|\eta|<2$ is calculated using:
\begin{equation}
\sigma = \frac{N_{\mathrm{data}} - N_{\mathrm{bkgd}}}{ \varepsilon_{\mathrm{cosmic}} \cdot \varepsilon_{\mathrm{FSR}} \cdot \varepsilon_{\mathrm{ee}} \cdot \varepsilon_{\mathrm{exc}} \cdot \eqlum }
\label{eq:csec}
\end{equation}
Considering that the exclusive process is derived from the
experimental observation (the sum of the exclusive and the
disassociation processes) we report both the total and the exclusive
cross sections.

The cross section for exclusive \mbox{$p\bar{p} \rightarrow p + e^+e^-
  + \bar{p}$} is measured to be $1.6
^{+0.5}_{-0.3}\mathrm{(stat)}\pm0.3\mathrm{(syst)}$~pb using
\mbox{$N_{\mathrm{bkgd}}=1.9\pm0.3$} events.  This agrees with the
theoretical cross section \mbox{$1.71\pm0.01$~pb} given by {\sc
  lpair}.  The probability of observing 16 events when $1.9\pm0.3$
events are expected is \mbox{$1.3 \times10^{-9}$}, equivalent to a
$5.5\sigma$ effect~\cite{pvalue}.

The measured cross section for \mbox{$\gamma\gamma \rightarrow
  e^+e^-$}, where all the proton and antiproton dissociation products
are contained within \mbox{$|\eta|>7.4$}, is measured to be \mbox{$1.8
  ^{+0.5}_{-0.3}\mathrm{(stat)}\pm0.2\mathrm{(syst)}~\mathrm{pb}$}
using \mbox{$N_{\mathrm{bkgd}}=0.3\pm0.1$} events (the sum of all
backgrounds except the dissociation).  This agrees with the
theoretical prediction of $1.9 \pm0.4{\rm(syst)}$~pb for this cross
section, determined using {\sc lpair}.  The uncertainty on this
prediction is larger than for the purely exclusive {\sc lpair}
prediction because assumptions about the hadronization of the
dissociating proton are made in this estimate.

In conclusion, we have observed 16 exclusive electron pair events in
CDF II, with \mbox{$|\eta|< 2.0$} and \mbox{$E_T>$5~GeV}, with a
background estimate of \mbox{$1.9\pm0.3$} events.  These events are
consistent in both their cross section and kinematic distributions
with \mbox{$p\bar{p} \rightarrow p + e^+e^- + \bar{p}$} through two
photon exchange \mbox{($\gamma\gamma \rightarrow e^+e^-$)}.  This is
the first time that exclusive two-photon processes have been observed
in hadron-hadron collisions, and implies that the LHC experiments can
rely on this process to calculate expectations for new physics and
luminosity measurements.

\section*{Acknowledgments}
We thank the Fermilab staff and the technical staffs of the
participating institutions for their vital contributions.  This work
was supported by the U.S. Department of Energy and National Science
Foundation; the Italian Istituto Nazionale di Fisica Nucleare; the
Ministry of Education, Culture, Sports, Science and Technology of
Japan; the Natural Sciences and Engineering Research Council of
Canada; the National Science Council of the Republic of China; the
Swiss National Science Foundation; the A.P. Sloan Foundation; the
Bundesministerium f\"ur Bildung und Forschung, Germany; the Korean
Science and Engineering Foundation and the Korean Research Foundation;
the Particle Physics and Astronomy Research Council and the Royal
Society, UK; the Russian Foundation for Basic Research; the Comisi\'on
Interministerial de Ciencia y Tecnolog\'{\i}a, Spain; in part by the
European Community's Human Potential Programme under contract
HPRN-CT-2002-00292; and the Academy of Finland.



\begin{thebibliography}{99}

\bibitem{bhabhabudnev} H. Bhabha, Proc. R. Soc. London Ser. A~{\bf 152}, 559 (1935);\\
 V. Budnev {\it et al.}, Phys. Lett. B~{\bf 39}, 526 (1972).
 
\bibitem{vermaseren} J. Vermaseren, Nucl. Phys.~{\bf B229}, 347 (1983).

\bibitem{newphyslhc} M.G.Albrow and A.Rostovtsev, hep-ph/0009336;\\
K. Piotrzkowski, Phys. Rev. D~{\bf 63}, 071502 (2001); \\
V. Khoze {\it et al.}, Eur. Phys. J. C~{\bf 23}, 311 (2002);\\
A. Kaidalov {\it et al.}, Eur. Phys. J. C~{\bf 31}, 387 (2003);\\  
A. Kaidalov {\it et al.}, Eur. Phys. J. C~{\bf 33}, 261 (2004);\\ 
V. Khoze {\it et al.}, Eur. Phys. J. C~{\bf 34}, 327 (2004).

\bibitem{fp420} CERN-LHCC-2005-025;LHCC-I-015 (www.fp420.com).


\bibitem{lhclumi} B. Caron and J. Pinfold, Helsinki 2000, Workshop on Forward Physics and Luminosity at the LHC;\\
V. Khoze {\it et al.}, Eur. Phys. J. C~{\bf 19}, 313 (2001)\\
A. Shamov and V. Telnov, Nucl. Instrum. Methods. A~{\bf 494}, 51 (2002); \\
D. Bocian and K. Piotrzkowski, Acta Phys. Polon. B~{\bf 35}, 2417 (2004).

\bibitem{cdfgg} A. Abulencia {\it et al.}, \emph{Evidence for Exclusive $\gamma\gamma$ Production in Hadron-Hadron Collisions},  to be submitted to Phys. Rev. Lett.

\bibitem{ep} A. Aktas {\it et al.}, Phys. Lett. B~{\bf 583}, 28 (2004);\\
A. Aktas {\it et al.}, Eur. Phys. J. C~{\bf 31}, 17 (2003).

\bibitem{aa} A. Belkacem {\it et al.}, Phys. Rev. A~{\bf 56}, 2806 (1997);\\
C. Vane {\it et al.}, Phys. Rev. A~{\bf 50}, 2313 (1997);\\
R. Baur {\it et al.}, Phys. Lett. B~{\bf 332}, 471 (1994);\\
J. Adams {\it et al.}, Phys. Rev. C~{\bf 70}, 031902 (2004).

\bibitem{isr} D. Antreasyan {\it et al.}, CERN-EP/80-82 (1980).

\bibitem{CDF} D. Acosta {\it et al.}, Phys. Rev. D~{\bf71}, 032001 (2005).

\bibitem{geometry} CDF II uses a cylindrical coordinate system with the $z$ coordinate along the proton beam direction, in which $\theta$ is the polar angle, $\phi$ is the azimuthal angle, and pseudorapidity \mbox{$\eta=-\ln \tan(\theta/2)$}.  \mbox{$E_T=E\sin\theta$} and \mbox{$p_T = p \sin \theta$} where $E$ is the energy measured by the calorimeter and $p$ the momentum measured in the tracking system.

\bibitem{ccal} L. Balka {\it et al.}, Nucl. Instrum. Methods A~{\bf 267}, 272 (1988);\\
S. Bertolucci {\it et al.}, Nucl. Instrum. Methods A~{\bf 267}, 301 (1988).

\bibitem{pem}  M. Albrow {\it et al.}, Nucl. Instrum. Methods A~{\bf 480}, 524 (2002).

\bibitem{emtiming} M. Goncharov {\it et al.}, FERMILAB-\-PUB-\-05-543-E (2005), accepted by Nucl. Instr. Methods, arXiv:physics/0512171.

\bibitem{miniplug} M. Gallinaro {\it et al.}, IEEE Trans. Nucl. Sci. {\bf 52}, 879 (2005).

\bibitem{COT} T. Affolder {\it et al.}, Nucl. Instrum. Methods A~{\bf 526}, 249 (2004).


\bibitem{cluster} The calorimeter is constructed of many separate segments, called towers.  A cluster is defined as a set of adjacent towers with an appreciable amount of energy deposited in them.

\bibitem{bunchlumi} Bunch luminosity is the instantaneous luminosity of each of the 36 individual bunch crossings at the Tevatron.

\bibitem{geant} GEANT, Detector description and simulation tool, CERN Program Library Long Writeup W5013 (1993).

\bibitem{cdfsim} E. Gerchtein and M. Paulini, Computing in High Energy and Nuclear Physics, preprint arXiv:physics-0306031, (2003).

\bibitem{yang} C.N. Yang, Phys. Rev. {\bf 77}, 242 (1950).

\bibitem{pythia} T. Sj\"ostrand {\it et al.}, Computer Phys. Commun. {\bf 135}, 238 (2001).  We use version 2.616.

\bibitem{grape} T. Abe, Comput. Phys. Commun. {\bf 136}, 126 (2001).

\bibitem{lpair-mbr} The {\sc lpair} simulation only describes the kinematics of the dissociating proton, so the Minimum Bias Rockefeller ({\sc mbr}) simulation~\cite{mbr} is used to simulate the products of the proton dissociation.  

\bibitem{mbr} F. Abe {\it et al.}, Phys. Rev. D~{\bf 50}, 5535 (1994).

\bibitem{pvalue} This probability is calculated from the sum of Poisson probabilities for seeing the observed number of events or more, convoluted with a Gaussian for the expected number of background events.

\end{thebibliography}
\end{document}